\documentclass{article}
\usepackage{spconf,amsmath,graphicx}
\usepackage{cite}
\usepackage{color,soul}

\usepackage{subfig}
\usepackage[colorlinks]{hyperref}
\hypersetup{
  colorlinks=true,
  citecolor=blue,
  linkcolor=red,
  urlcolor=blue}
\usepackage{setspace}

\usepackage{etoolbox}
\patchcmd{\thebibliography}
  {\settowidth}
  {\setlength{\itemsep}{0pt plus 0.pt}\settowidth}
  {}{}

\title{A Comparative Evaluation of Temporal Pooling Methods for Blind Video Quality Assessment}
\name{Zhengzhong~Tu$^{1\star}$\thanks{$^\star$Equal contribution},~Chia-Ju~Chen$^{1\star}$,~Li-Heng~Chen$^1$,~Neil~Birkbeck$^2$,~Balu~Adsumilli$^2$,~and~Alan~C.~Bovik$^1$}
\address{$^1$The~University~of~Texas~at~Austin,~$^2$YouTube~Media~Algorithms~Team,~Google~Inc.}
\begin{document}
%
\maketitle
\begin{abstract}
Many objective video quality assessment (VQA) algorithms include a key step of temporal pooling of frame-level quality scores. However, less attention has been paid to studying the relative efficiencies of different pooling methods on no-reference (blind) VQA. Here we conduct a large-scale comparative evaluation to assess the capabilities and limitations of multiple temporal pooling strategies on blind VQA of user-generated videos. The study yields insights and general guidance regarding the application and selection of temporal pooling models. In addition, we also propose an ensemble pooling model built on top of high-performing temporal pooling models.
Our experimental results demonstrate the relative efficacies of the evaluated temporal pooling models, using several popular VQA algorithms, and evaluated on two recent large-scale natural video quality databases. In addition to the new ensemble model, we provide a general recipe for applying temporal pooling of frame-based quality predictions.
\end{abstract}
\section{Introduction}
\label{sec:intro}
Video quality assessment (VQA) models have been widely studied \cite{seshadrinathan2010study} as an increasingly important toolset used by the streaming and social media industries. While full-reference (FR) VQA research is gradually maturing and several algorithms \cite{wang2004image, li2016toward} are quite widely deployed, recent attention has shifted more towards creating better no-reference (NR) VQA models that can be used to predict and monitor the quality of authentically distorted user-generated content (UGC) videos. UGC videos, which are typically created by amateur videographers, often suffer from unsatisfactory perceptual quality, arising from imperfect capture devices, uncertain shooting skills, a variety of possible content processes, as well as compression and streaming distortions. In this regard, predicting UGC video quality is much more challenging than assessing the quality of synthetically distorted videos in traditional video databases. UGC distortions are more diverse, complicated, commingled, and no ``pristine'' reference is available. 

Many researchers have studied and proposed possible solutions to the NR VQA problem \cite{mittal2012no, saad2014blind,xu2014no, mittal2015completely, ghadiyaram2017perceptual, varga2019no, li2019quality}, among which a simple but reasonably effective strategy is to compute frame-level quality scores, e.g., as generated by image quality assessment (IQA) models, then to express the evolution or relative importance over time by applying temporal pooling on the frame-level quality scores. Simple temporal average pooling is a widely used scheme to augment both FR \cite{seshadrinathan2009motion, bampis2017speed, vu2014vis3} and NR VQA models \cite{mittal2015completely, saad2014blind,varga2019no}. Other kinds of pooling that are used include harmonic mean \cite{li2018vmaf}, Minkowski mean \cite{rimac2009influence, seufert2013pool}, percentile pooling \cite{moorthy2009visual,chen2016perceptual}, and adaptively weighted sums \cite{park2012video}. More sophisticated pooling strategies have considered memory effects, such as primacy, recency \cite{bampis2017study, rimac2009influence, seufert2013pool}, and hysteresis \cite{seshadrinathan2011temporal, xu2014no, li2019quality, choi2018video}. The general applicability of these pooling models, however, has not so far been deeply validated in the general context of NR VQA models for real-world UGC videos, though a few more directed studies have been conducted \cite{rimac2009influence, seufert2013pool}. To date, no comprehensive studies have been conducted to establish the added values of the spectrum of available VQA pooling schemes.

Here we seek to help fill this gap by conducting a systematic evaluation of popular temporal pooling algorithms, as applied to leading NR IQA models on recently developed large scale UGC video quality databases. We assessed the benefits, generalizability, and stability of these pooling mechanisms. Our aim is to identify statistically verifiable pooling approaches that can be applied on top of future state-of-the-art IQA models to further produce consistently better predictions of video quality. We also propose an ensemble approach, wherein multiple pooling models are aggregated to deliver better retrospective quality prediction. Our experimental results demonstrate that the proposed ensemble pooling method reveals robustness among the top-performing models.

The rest of this paper is structured as follows. Section \ref{sec:related_work} summarizes previous related literature, while Section \ref{sec:Ensemble} describes details of the evaluated and proposed pooling algorithms. Experimental results and analysis are presented in Section \ref{sec:experiments}, and finally, we conclude the paper in Section \ref{sec:conclusion}.

\section{Related Work}
\label{sec:related_work}

A variety of methods for spatial pooling of ``quality-aware'' features have been proposed and studied in \cite{engelke2011visual, moorthy2009visual, wang2010information}, yet less effort has been applied to the study on temporal pooling methods for NR VQA. The most related works to that reported here are the comparative evaluations of temporal pooling on short video clips \cite{rimac2009influence}, and on longer adaptive streaming videos \cite{seufert2013pool}. They have collectively included various pooling methods combined with several objective frame-level quality predictors, evaluated on different subjective databases. Among the studied temporal pooling methods are: simple averaging, percentile pooling \cite{moorthy2009visual}, Minkowski pooling \cite{rimac2009influence}, harmonic mean pooling \cite{li2018vmaf}, and the more complex VQPooling scheme \cite{park2012video}, which adaptively emphasizes the worst scores along the time dimension, wherein frame-level scores are clustered into two groups (low quality and high quality), then combined into a single score by upweighting low-quality scores. Methods like percentile and VQPooling are predicated by the accepted notion that quality judgments are heavily influenced by the worst parts of a video.

Another cognitive aspect relevant to temporal visual pooling is the serial-position effect (or memory effect) hypothesis \cite{murdock1962serial}. Primacy and recency are two common effects that have been investigated in numerous video quality of experience (QoE) studies \cite{bampis2017study, ghadiyaram2018learning, nguyen2019modeling}, but are less studied in regard to their influence on the blind quality prediction of UGC video clips. Another popular temporal memory modeling approach is hysteresis pooling \cite{seshadrinathan2011temporal}, which has been justified in several video quality modeling papers \cite{xu2014no, li2019quality, choi2018video}. The hysteresis model assumes that while subjective judgments drop sharply with event of poor video quality, they only recover slowly with subsequent improved video quality.



\section{Evaluating Temporal Pooling Methods}
\label{sec:Ensemble}

We propose a comprehensive evaluation framework to study the influence of temporal pooling algorithms on the performances of objective video quality models.
Suppose a video has $N$ frames $\{F_1,F_2,...,F_N\}$ processed by any NR IQA models that produces frame-level (time-varying) quality predictions $\{q_1,q_2,...q_N\}$. The per-frame quality scores are temporally combined by a temporal pooling function $\mathcal{F}(\cdot)$ to obtain a final quality prediction: $Q_\mathrm{FINAL}=\mathcal{F}(q_1,q_2,...,q_N)$. 

\subsection{Frame Quality Prediction}
\label{ssec:frame-quality}

Frame-level quality scores can be predicted by any NR IQA, such as BRISQUE \cite{mittal2012no}, NIQE \cite{mittal2012making}, FRIQUEE \cite{ghadiyaram2017perceptual} or even models implemented as deep learning networks \cite{kim2017deep}.

\subsection{Temporal Pooling Models}
\label{ssec:temporal-pool-prox}

Once frame-level quality scores $\{q_1,q_2,...q_N\}$ are obtained, a variety of ways have been proposed to summarize the time-varying quality scores into a single overall video quality judgment. A variety of human factors have been explored in this context, including visual perception \cite{de2013model, chen2014modeling}, memory effects \cite{bampis2017study, ghadiyaram2018learning, seshadrinathan2011temporal}, and video content \cite{ghadiyaram2018learning, mirkovic2014evaluating, li2019quality}. Here we model and study a collection of factors that express aspects of temporal quality perception, as candidates for deriving final quality predictions on UGC videos. Specifically, we study the following listed in approximate order of increasing complexity and abstraction:

\noindent\textbf{Arithmetic Mean}: The sample mean of frame-level scores is the most widely used method: 
\begin{equation}
\label{eq:am}   
Q=\frac{1}{N}\sum_{n=1}^N q_n.
\end{equation}

\noindent\textbf{Harmonic Mean}: The harmonic mean has been observed to emphasize the impact of low-quality frames \cite{li2018vmaf}:
\begin{equation}
\label{eq:hm}   
Q=\left(\frac{1}{N}\sum_{n=1}^N q_n^{-1}\right)^{-1}.
\end{equation}

\noindent\textbf{Geometric Mean}: The third Pythagorean mean (geometric) expresses the central tendency of the quality scores by the product of their values:
\begin{equation}
\label{eq:gm}   
Q=\left(\prod_{n=1}^N q_n\right)^{1/N}.
\end{equation}

\noindent\textbf{Minkowski Mean}: The $L_p$ Minkowski summation \cite{rimac2009influence, seufert2013pool} of time-varying quality is defined as:
\begin{equation}
\label{eq:mm}   
Q=\left(\frac{1}{N}\sum_{n=1}^N q_n^{p}\right)^{1/p}.
\end{equation}

\noindent\textbf{Percentile}: The idea of percentile pooling is based on observed phenomenon that perceptual quality is heavily affected by the ``worst'' parts of the content. Many prior works have studied and justified (or challenged) percentile pooling \cite{moorthy2009visual, chen2016perceptual, seufert2013pool, rimac2009influence, bampis2017study}. The $k$-th percentile pooling is expressed:
\begin{equation}
\label{eq:perc}   
Q=\frac{1}{|P_{\downarrow k\%}|}\sum_{n\in P_{\downarrow k\%}} q_n.
\end{equation}

\noindent\textbf{VQPooling}: VQPooling is an adaptive spatial and temporal pooling strategy proposed in \cite{park2012video}. Here we only study the temporal pooling part, wherein the quality scores of all frames are classified into two groups composed of higher and lower quality, using $k$-means clustering. The two groups, dubbed $G_L$ and $G_H$, are then combined to obtain an overall quality prediction on the entire video sequence:
\begin{equation}
\label{eq:vqp}   
Q=\frac{\sum_{n\in G_L} q_n+w\cdot \sum_{n\in G_H} q_n}{|G_L|+w\cdot |G_H|},
\end{equation}
where $|G_L|$ and $|G_H|$ denote the cardinality of $G_L$ and $G_H$, while the weight $w$ is defined as the ratio between the scores in $G_L$ and $G_H$:
\begin{equation}
\label{eq:}
w=\left(1-\frac{M_L}{M_H}\right)^2,
\end{equation}
where $M_L$ and $M_H$ are the average value of the quality scores in set $G_L$ and $G_H$, respectively.

\noindent\textbf{Temporal Variation}: The approach of \cite{ninassi2009considering} considers the temporal changes of spatial distortions over time and proposes short-term and long-term spatiotemporal pooling mechanisms to account for quality changes. Here we only utilize the temporal variation terms in our study:
\begin{equation}
\label{eq:variation}
Q=\frac{1}{|P_{\uparrow k\%}|}\sum_{n\in P_{\uparrow k\%}} |\nabla q_n|, 
\end{equation}
where $|\nabla q_n|$ is the absolute gradient at time $n$, and $Q$ pools the largest $k\%$ of the per-frame gradients of quality values.

\noindent\textbf{Primacy Effect}: The primacy effect describes the tendency of human viewers to recall the earliest portion of a video when providing overall evaluations \cite{murdock1962serial}. One way of capturing primacy is as an exponentially decreasing weighted sum. Define
\begin{equation}
\label{eq:primacy}
Q=\sum_{n=1}^N w_n  q_n,
\end{equation}
where
\begin{equation}
\label{eq:w_p}
w_n=\frac{\exp{(-\alpha_p n)}}{\sum_{k=0}^L\exp{(-\alpha_p k})},\ 0\leq n\leq L.
\end{equation}


\noindent\textbf{Recency Effect}: The recency effect is another well-established behavioral and memory effect, whereby, in this context, video quality is very strongly influenced by a viewer's most recently percieved visual impression \cite{murdock1962serial}. The recency effect can also be characterized as an exponential weighted sum (Eq. (\ref{eq:primacy})), but with a different weighting:
\begin{equation}
\label{eq:w_r}
w_n=\frac{\exp{(-\alpha_r(L-n))}}{\sum_{k=0}^L\exp{(-\alpha_r(L-k)})},\ 0\leq n\leq L,
\end{equation}
where $\alpha_p$ in (\ref{eq:w_p}) and $\alpha_r$ in (\ref{eq:w_r}) can be used to tune the relative intensity of these two memory effects. 

\noindent\textbf{Temporal Hysteresis}: This approach was inspired by the hysteresis effect observed in human judgments of time-varying video quality \cite{seshadrinathan2011temporal}, which is closely related to, but not the same as the recency effect. The hysteresis measurement can be formulated as follows. Let $q_n,\ n=1,2,...N$ be the time-varying frame quality scores. The memory of past quality $l_n$ at the $n$-th frame is expressed as the minimum quality scores over the previous frames:
\begin{equation}
\label{eq:memory}
l_n=\left\{
\begin{array}{ll}
    q_n, & n=1 \\
    \min\limits_{k\in \mathcal{K}_{prev}}\{q_k\}, & n>1,
\end{array}
\right.
\end{equation}
where $\mathcal{K}_{prev}=\{\max\{1,n-\tau\},...,n-2,n-1\}$ indexes the previous $\tau$ frames. The current video quality $m_n$ is expressed as a weighted sum of ordered \cite{longbotham1989theory} frame-level qualities:
\begin{equation}
\label{eq:current}
\boldsymbol{v}=sort(\{q_k\}),\ k\in \mathcal{K}_{next},
\end{equation}
\begin{equation}
\label{eq:current2}
m_n=\sum_{j=1}^J v_j w_j,\ J=|\mathcal{K}_{next}|,
\end{equation}
where $\mathcal{K}_{next}=\{n,n+1,...,\min\{n+\tau,N\}\}$ indexes the next $\tau$ frames and $\{w_j\}$ is the descending half of a Gaussian weighting function. Linearly combining the memory and the current quality components in (\ref{eq:memory}) and (\ref{eq:current2}) yields time-varying scores that capture the hysteresis effect. The pooled video quality $Q$ is computed as the global temporal average of the time-varying hysteresis-transformed predictions:
\begin{equation}
\label{eq:linear}
q^\prime_{n}=\alpha m_n+(1-\alpha) l_n,
\end{equation}
\begin{equation}
\label{eq:avg}
Q=\frac{1}{N}\sum_{n=1}^N q^\prime_{n},
\end{equation}
where $\alpha$ adjusts the contributions of these two elements.


\begin{table*}[!t]
\footnotesize	
\setlength{\tabcolsep}{3.05pt}
\caption{Performance comparison of temporal pooling methods as evaluated on KoNViD-1k \cite{hosu2017konstanz} and LIVE-VQC \cite{sinno2018large}. Each cell shows the median evaluation results formatted as SRCC\! /\! PLCC. The three best results along each column are \textbf{boldfaced}.}
\label{table:1}
\centering
\begin{tabular}{lccccccccccc}
\hline\hline
Database & \multicolumn{5}{c}{KoNViD-1k} & & \multicolumn{5}{c}{LIVE-VQC} \\ \cline{2-6}\cline{8-12}
Pool/Model & NIQE & BRISQUE &  GMLOG &  HIGRADE & CORNIA & & NIQE & BRISQUE &  GMLOG &  HIGRADE & CORNIA \\
\hline
Mean & 0.552\! /\! \textbf{0.560} & 0.673\! /\! 0.676 & 0.662\! /\! 0.671  & 0.690\! /\! 0.696 & \textbf{0.749}\! /\! \textbf{0.764} & & 0.600\! /\! 0.631 & 0.597\! /\! 0.632 & 0.575\! /\! 0.618 & 0.532\! /\! 0.570  & 0.694\! /\! 0.743  \\
Median & 0.543\! /\! 0.554 & 0.667\! /\! 0.670 & 0.657\! /\! 0.666 & 0.680\! /\! 0.689 & \textbf{0.750}\! /\! \textbf{0.760} & & 0.584\! /\! 0.618  & 0.577\! /\! 0.619 & 0.558\! /\! 0.602 & 0.521\! /\! 0.559 & 0.687\! /\! 0.744 \\
Harmonic & 0.550\! /\! \textbf{0.560} & \textbf{0.674}\! /\! 0.676  & 0.667\! /\! 0.674 & 0.693\! /\! 0.699 & 0.696\! /\! 0.696 & & 0.607\! /\! 0.637  & 0.605\! /\! 0.636 & 0.585\! /\! 0.620 &   0.537\! /\! 0.575 & \textbf{0.709}\! /\! 0.737 \\
Geometric & 0.551\! /\! \textbf{0.560} &  \textbf{0.676}\! /\! \textbf{0.679} & 0.666\! /\! 0.673 & 0.692\! /\! 0.698 & 0.747\! /\! \textbf{0.760} & & 0.604\! /\! 0.634 & 0.600\! /\! 0.631 & 0.578\! /\! 0.617 & 0.537\! /\! 0.573 & 0.698\! /\! \textbf{0.746} \\

Minkowski & 0.552\! /\! 0.559 & 0.672\! /\! 0.676 & 0.661\! /\! 0.670 & 0.689\! /\! 0.695 & 0.736\! /\! 0.746 & & 0.597\! /\! 0.628 & 0.596\! /\! 0.630 & 0.574\! /\! 0.615 & 0.538\! /\! 0.569 & 0.688\! /\! 0.739 \\
Percentile & 0.545\! /\! 0.547 & 0.655\! /\! 0.647  & \textbf{0.674}\! /\! \textbf{0.678} & 0.685\! /\! 0.687 & 0.696\! /\! 0.700 & & \textbf{0.630}\! /\! 0.634  & \textbf{0.629}\! /\! \textbf{0.647} &  \textbf{0.606}\! /\! \textbf{0.627} & \textbf{0.586}\! /\! \textbf{0.610} & \textbf{0.712}\! /\! 0.744 \\
VQPooling & 0.549\! /\! 0.554 & 0.670\! /\! 0.665 & \textbf{0.672}\! /\! 0.674 & \textbf{0.698}\! /\! \textbf{0.701} & 0.743\! /\! 0.758 & & \textbf{0.628}\! /\! \textbf{0.644} & \textbf{0.617}\! /\! \textbf{0.658} &  \textbf{0.605}\! /\! \textbf{0.633} & 0.563\! /\! 0.597 & 0.700\! /\! \textbf{0.753} \\

Variation & 0.347\! /\! 0.328 & 0.348\! /\! 0.338 & 0.509\! /\! 0.511 & 0.434\! /\! 0.444 & 0.240\! /\! 0.303 & & 0.507\! /\! 0.476 & 0.470\! /\! 0.463 & 0.495\! /\! 0.488 & 0.474\! /\! 0.482 & 0.567\! /\! 0.609  \\
Primacy & 0.541\! /\! 0.552 & 0.668\! /\! 0.671 & 0.647\! /\! 0.653 & 0.684\! /\! 0.690 & 0.726\! /\! 0.741 & & 0.601\! /\! 0.631 & 0.573\! /\! 0.627 & 0.575\! /\! 0.613 & 0.535\! /\! 0.561 & 0.684\! /\! 0.737 \\
Recency & \textbf{0.553}\! /\! 0.558 & 0.670\! /\! 0.667 & 0.660\! /\! 0.667 & 0.690\! /\! 0.694 & 0.745\! /\! 0.754 &  & 0.584\! /\! 0.615 & 0.586\! /\! 0.626 & 0.561\! /\! 0.599 & 0.518\! /\! 0.555 & 0.670\! /\! 0.729 \\

Hysteresis &  \textbf{0.563}\! /\! \textbf{0.569} & \textbf{0.684}\! /\! \textbf{0.681} & \textbf{0.681}\! /\! \textbf{0.684} & \textbf{0.703}\! /\! \textbf{0.707} & 0.732\! /\! 0.735 & & 0.621\! /\! \textbf{0.638} & \textbf{0.621}\! /\! \textbf{0.650} &  \textbf{0.600}\! /\! \textbf{0.629} & \textbf{0.570}\! /\! \textbf{0.595} & \textbf{0.711}\! /\! \textbf{0.756} \\
EPooling & \textbf{0.572}\! /\! \textbf{0.579} &  0.670\! /\! \textbf{0.679} & 0.670\! /\! \textbf{0.676} & \textbf{0.698}\! /\! \textbf{0.704}  & \textbf{0.749}\! /\! \textbf{0.762} & & \textbf{0.623}\! /\! \textbf{0.645} & \textbf{0.617}\! /\! 0.646  & \textbf{0.605}\! /\! 0.623  & \textbf{0.582}\! /\! \textbf{0.601} & 0.705\! /\! 0.743 \\


\hline\hline
\end{tabular}
\end{table*}

\subsection{Ensemble Temporal Pooling}
\label{ssec:ensemble-pool}

We have just described a diverse set of temporal pooling mechanisms, each either heuristically, statistically defined, or motivated by psychovisual reasoning. As might be expected, and as we shall show, the performances of these methods differ, and also vary on different datasets. Given that these methods likely capture different aspects of perceptual pooling, ensemble learning is a direct way to combine them towards creating a more reliable and generic quality predictor. We denote this ensemble-based temporal pooling as \textbf{EPooling}. Similar concepts of model fusion/ensemble have been successfully utilized on the IQA/VQA problems \cite{bampis2018spatiotemporal, pei2015image, li2016toward}. 

Suppose the quality scores delivered by a set of pooling methods are denoted $Q_i,\ i=1,...,I$, where $I$ is the number of input model predictions. Then train an ensemble regressor to fuse the multiple predicted labels into a single final score:
\begin{equation}
\label{eq:fusion}
Q_{\mathrm{EPooling}}=\mathcal{F}(\boldsymbol{Q}),\ \boldsymbol{Q}=\{Q_i\},\ i=1,2,...,I,
\end{equation}
where $\boldsymbol{Q}$ is the quality vector stacked from multiple singly pooled scores, and $\mathcal{F}$ is the learned regression function that maps the proxy quality vector to a final quality prediction $Q_{\mathrm{EPooling}}$. Here we empirically chose Mean, VQPooling, and Hysteresis, as the three input prediction models after coarse preliminary feature analysis. Further improvements may be achieved by applying finer feature selection techniques.

\section{Experiments}
\label{sec:experiments}

\subsection{Experimental Setup}
\label{ssec:exp-setup}
We selected five popular NR IQA models: NIQE \cite{mittal2012making}, BRISQUE \cite{mittal2012no}, GM-LOG \cite{xue2014blind}, HIGRADE \cite{kundu2017no}, and CORNIA \cite{ye2012unsupervised}, as frame-level quality predictors, and evaluated the temporal pooling methods on two recent large scale UGC VQA databases: KoNViD-1k \cite{hosu2017konstanz} and LIVE-VQC \cite{sinno2018large}. When defining the parametric temporal pooling models, we used $p=2\ (\mathcal{L}^2)$ for Minkowski, $k=10\%$ for percentile, $(L,\alpha_p,\alpha_r)=(180,0.01,0.01)$ for primacy and recency, and $(\tau,\alpha)=(60,0.8)$ for Temporal Hysteresis, as recommended in the originating works. We randomly split the evaluation dataset into $80\%$-$20\%$ portions for training and testing, respectively, over $100$ trials and report the overall median performance on the testing set. We only conducted $20$ iterations for CORNIA due to its high training complexity. Within each split iteration, EPooling requires two phases of training -- first, to train the mapping from the IQA feature vector to frame-level quality predictions (meaning predicted MOS), then, to learn the regression function that fuses the temporally pooled predictions to obtain the final quality result. Both phases are conducted on the training set. We used a support vector regression (SVR) as the learning model for both training stages, employing cross-validation and $3\times3$ grid-search for the SVR parameter selection. As performance metrics, we used the Spearman rank-order correlation coefficient (SRCC) calculated between the ground truth MOS and the predicted scores to measure the prediction monotonicity of the models, and the Pearson linear correlation coefficient (PLCC) (computed after logistic mapping) to measure the degree of linear correlation against MOS.

\subsection{Results and Recipe}
\label{ssec:res-rec}
The performance results are shown in Table \ref{table:1} on the KoNViD-1k \cite{hosu2017konstanz} and LIVE-VQC \cite{sinno2018large}, respectively. On KoNViD-1k, none of the sophisticated pooling algorithms were observed to significantly outperform the sample mean of temporal video quality scores. While an average gain of $\sim$\! $0.01$ in SRCC/PLCC was achieved using Hysteresis pooling, the three classical Pythagorean means performed quite well despite their simplicity and computational efficiency. When tested on LIVE-VQC \cite{sinno2018large}, however, we have observed a $\sim$\! $0.03$ performance average gain when employing perceptual importance pooling methods like percentile \cite{moorthy2009visual}, VQPooling \cite{park2012video}, and Hysteresis \cite{seshadrinathan2011temporal}, regardless of which NR IQA model was used. It is likely that the memory-related effects, primacy and recency, would play a more important role on longer videos (usually minutes long), as shown in \cite{bampis2017study, ghadiyaram2018learning}, but they did not contribute much on the short duration videos (8-10 seconds) in these datasets. Our proposed ensemble method of pooling achieved consistently competitive outcomes on both datasets.

These performance results yet reveal different trends on the two databases: KoNViD-1k yielded similar results among most of the competing pooling approaches, whereas on LIVE-VQC, Percentile, VQPooling, Hysteresis, and the ensemble enhancement, EPooling, generated the best scores. Towards understanding this, we observe that LIVE-VQC contains videos with more camera motion, hence more temporal variation than those in KoNViD-1k. It is possible that LIVE-VQC contains a larger range of perceived time-varying qualities scores, while temporal quality variations in KoNViD-1k adhere more closely to the mean quality level. Recalling the aforementioned hypothesis that perceptual quality is heavily affected by the worst portions of a video, our experimental results promote this assumption. In conclusion, our suggested recipe for incorporating temporal pooling into the design of NR VQA models strongly depends on video content -- for videos containing more motion or temporal quality variations, pooling strategies that more heavily weight low quality events are recommended. In situations where the quality variations are low, or contain less motion, traditional statistical mean predictions may be adequate.


\section{Conclusion}
\label{sec:conclusion}

We conducted a benchmark study on the added value of integrating temporal pooling into blind video quality assessment for user-generated video content. We found that the efficacy of temporal pooling is content-dependent, but an ensemble approach can further improve quality prediction performance on a difficult problem that is only incompletely understood.




\section{references}
\label{sec:ref}

\bibliographystyle{IEEEtran}
\
\footnotesize\bibliography{refs}

\begin{thebibliography}{10}
\providecommand{\url}[1]{#1}
\csname url@samestyle\endcsname
\providecommand{\newblock}{\relax}
\providecommand{\bibinfo}[2]{#2}
\providecommand{\BIBentrySTDinterwordspacing}{\spaceskip=0pt\relax}
\providecommand{\BIBentryALTinterwordstretchfactor}{4}
\providecommand{\BIBentryALTinterwordspacing}{\spaceskip=\fontdimen2\font plus
\BIBentryALTinterwordstretchfactor\fontdimen3\font minus
  \fontdimen4\font\relax}
\providecommand{\BIBforeignlanguage}[2]{{%
\expandafter\ifx\csname l@#1\endcsname\relax
\typeout{** WARNING: IEEEtran.bst: No hyphenation pattern has been}%
\typeout{** loaded for the language `#1'. Using the pattern for}%
\typeout{** the default language instead.}%
\else
\language=\csname l@#1\endcsname
\fi
#2}}
\providecommand{\BIBdecl}{\relax}
\BIBdecl

\bibitem{seshadrinathan2010study}
K.~Seshadrinathan, R.~Soundararajan, A.~C. Bovik, and L.~K. Cormack, ``Study of
  subjective and objective quality assessment of video,'' \emph{IEEE Trans.
  Image Process.}, vol.~19, no.~6, pp. 1427--1441, 2010.

\bibitem{wang2004image}
Z.~Wang, A.~C. Bovik, H.~R. Sheikh, and E.~P. Simoncelli, ``Image quality
  assessment: from error visibility to structural similarity,'' \emph{IEEE
  Trans. Image Process.}, vol.~13, no.~4, pp. 600--612, 2004.

\bibitem{li2016toward}
Z.~Li, A.~Aaron, I.~Katsavounidis, A.~Moorthy, and M.~Manohara, ``Toward a
  practical perceptual video quality metric,'' \emph{The Netflix Tech Blog},
  vol.~6, 2016.

\bibitem{mittal2012no}
A.~Mittal, A.~K. Moorthy, and A.~C. Bovik, ``No-reference image quality
  assessment in the spatial domain,'' \emph{IEEE Trans. Image Process.},
  vol.~21, no.~12, pp. 4695--4708, 2012.

\bibitem{saad2014blind}
M.~A. Saad, A.~C. Bovik, and C.~Charrier, ``Blind prediction of natural video
  quality,'' \emph{IEEE Trans. Image Process.}, vol.~23, no.~3, pp. 1352--1365,
  2014.

\bibitem{xu2014no}
J.~Xu, P.~Ye, Y.~Liu, and D.~Doermann, ``No-reference video quality assessment
  via feature learning,'' in \emph{Proc. IEEE Int. Conf. Image Process.}, 2014,
  pp. 491--495.

\bibitem{mittal2015completely}
A.~Mittal, M.~A. Saad, and A.~C. Bovik, ``A completely blind video integrity
  oracle,'' \emph{IEEE Trans. Image Process.}, vol.~25, no.~1, pp. 289--300,
  2015.

\bibitem{ghadiyaram2017perceptual}
D.~Ghadiyaram and A.~C. Bovik, ``Perceptual quality prediction on authentically
  distorted images using a bag of features approach,'' \emph{J. Vis.}, vol.~17,
  no.~1, pp. 32--32, 2017.

\bibitem{varga2019no}
D.~Varga, ``No-reference video quality assessment based on the temporal pooling
  of deep features,'' \emph{Neural Process. Lett.}, pp. 1--14, 2019.

\bibitem{li2019quality}
D.~Li, T.~Jiang, and M.~Jiang, ``Quality assessment of in-the-wild videos,'' in
  \emph{Proc. ACM Int. Conf. Multimedia}, 2019, pp. 2351--2359.

\bibitem{seshadrinathan2009motion}
K.~Seshadrinathan and A.~C. Bovik, ``Motion tuned spatio-temporal quality
  assessment of natural videos,'' \emph{IEEE Trans. Image Process.}, vol.~19,
  no.~2, pp. 335--350, 2009.

\bibitem{bampis2017speed}
C.~G. Bampis, P.~Gupta, R.~Soundararajan, and A.~C. Bovik, ``{SpEED-QA}:
  spatial efficient entropic differencing for image and video quality,''
  \emph{IEEE Signal Process. Lett.}, vol.~24, no.~9, pp. 1333--1337, 2017.

\bibitem{vu2014vis3}
P.~V. Vu and D.~M. Chandler, ``{ViS3}: An algorithm for video quality
  assessment via analysis of spatial and spatiotemporal slices,'' \emph{J.
  Electron. Imag.}, vol.~23, no.~1, p. 013016, 2014.

\bibitem{li2018vmaf}
Z.~Li, C.~Bampis, J.~Novak, A.~Aaron, K.~Swanson, A.~Moorthy, and J.~Cock,
  ``{VMAF}: The journey continues,'' \emph{Netflix Technology Blog}, 2018.

\bibitem{rimac2009influence}
S.~Rimac-Drlje, M.~Vranjes, and D.~Zagar, ``Influence of temporal pooling
  method on the objective video quality evaluation,'' in \emph{Proc. IEEE Int.
  Symp. Broadband Multimedia Syst. Broadcast.}, 2009, pp. 1--5.

\bibitem{seufert2013pool}
M.~Seufert, M.~Slanina, S.~Egger, and M.~Kottkamp, ``{``To pool or not to
  pool''}: a comparison of temporal pooling methods for http adaptive video
  streaming,'' in \emph{Proc. Int. Workshop Qual. Multimedia Exper. (QoMEX)},
  2013, pp. 52--57.

\bibitem{moorthy2009visual}
A.~K. Moorthy and A.~C. Bovik, ``Visual importance pooling for image quality
  assessment,'' \emph{IEEE J. Sel. Topics Signal Process.}, vol.~3, no.~2, pp.
  193--201, 2009.

\bibitem{chen2016perceptual}
C.~Chen, M.~Izadi, and A.~Kokaram, ``A perceptual quality metric for videos
  distorted by spatially correlated noise,'' in \emph{Proc. ACM Int. Conf.
  Multimedia}, 2016, pp. 1277--1285.

\bibitem{park2012video}
J.~Park, K.~Seshadrinathan, S.~Lee, and A.~C. Bovik, ``Video quality pooling
  adaptive to perceptual distortion severity,'' \emph{IEEE Trans. Image
  Process.}, vol.~22, no.~2, pp. 610--620, 2012.

\bibitem{bampis2017study}
C.~G. Bampis, Z.~Li, A.~K. Moorthy, I.~Katsavounidis, A.~Aaron, and A.~C.
  Bovik, ``Study of temporal effects on subjective video quality of
  experience,'' \emph{IEEE Trans. Image Process.}, vol.~26, no.~11, pp.
  5217--5231, 2017.

\bibitem{seshadrinathan2011temporal}
K.~Seshadrinathan and A.~C. Bovik, ``Temporal hysteresis model of time varying
  subjective video quality,'' in \emph{Proc. IEEE Int. Conf. Acoust., Speech,
  Signal Process.}, 2011, pp. 1153--1156.

\bibitem{choi2018video}
L.~K. Choi and A.~C. Bovik, ``Video quality assessment accounting for temporal
  visual masking of local flicker,'' \emph{Signal Process.: Image Commun.},
  vol.~67, pp. 182--198, 2018.

\bibitem{engelke2011visual}
U.~Engelke, H.~Kaprykowsky, H.-J. Zepernick, and P.~Ndjiki-Nya, ``Visual
  attention in quality assessment,'' \emph{IEEE Signal Process. Mag.}, vol.~28,
  no.~6, pp. 50--59, 2011.

\bibitem{wang2010information}
Z.~Wang and Q.~Li, ``Information content weighting for perceptual image quality
  assessment,'' \emph{IEEE Trans. Image Process.}, vol.~20, no.~5, pp.
  1185--1198, 2010.

\bibitem{murdock1962serial}
B.~B. Murdock~Jr, ``The serial position effect of free recall,'' \emph{J. Exp.
  Psychol.}, vol.~64, no.~5, p. 482, 1962.

\bibitem{ghadiyaram2018learning}
D.~Ghadiyaram, J.~Pan, and A.~C. Bovik, ``Learning a continuous-time streaming
  video {QoE} model,'' \emph{IEEE Trans. Image Process.}, vol.~27, no.~5, pp.
  2257--2271, 2018.

\bibitem{nguyen2019modeling}
T.~Nguyen~Duc, C.~Minh~Tran, P.~X. Tan, and E.~Kamioka, ``Modeling of
  cumulative {QoE} in on-demand video services: Role of memory effect and
  degree of interest,'' \emph{Future Internet}, vol.~11, no.~8, p. 171, 2019.

\bibitem{mittal2012making}
A.~Mittal, R.~Soundararajan, and A.~C. Bovik, ``Making a “completely blind”
  image quality analyzer,'' \emph{IEEE Signal Process. Lett.}, vol.~20, no.~3,
  pp. 209--212, 2012.

\bibitem{kim2017deep}
J.~Kim, H.~Zeng, D.~Ghadiyaram, S.~Lee, L.~Zhang, and A.~C. Bovik, ``Deep
  convolutional neural models for picture-quality prediction: Challenges and
  solutions to data-driven image quality assessment,'' \emph{IEEE Signal
  Process. Mag.}, vol.~34, no.~6, pp. 130--141, 2017.

\bibitem{de2013model}
J.~De~Vriendt, D.~De~Vleeschauwer, and D.~Robinson, ``Model for estimating
  {QoE} of video delivered using {HTTP} adaptive streaming,'' in \emph{Proc.
  IFIP/IEEE Int Symp. Integr. Netw. Manag.}, 2013, pp. 1288--1293.

\bibitem{chen2014modeling}
C.~Chen, L.~K. Choi, G.~De~Veciana, C.~Caramanis, R.~W. Heath, and A.~C. Bovik,
  ``Modeling the time-varying subjective quality of {HTTP} video streams with
  rate adaptations,'' \emph{IEEE Trans. Image Process.}, vol.~23, no.~5, pp.
  2206--2221, 2014.

\bibitem{mirkovic2014evaluating}
M.~Mirkovic, P.~Vrgovic, D.~Culibrk, D.~Stefanovic, and A.~Anderla,
  ``Evaluating the role of content in subjective video quality assessment,''
  \emph{Sci. World J.}, vol. 2014, 2014.

\bibitem{ninassi2009considering}
A.~Ninassi, O.~Le~Meur, P.~Le~Callet, and D.~Barba, ``Considering temporal
  variations of spatial visual distortions in video quality assessment,''
  \emph{IEEE J. Sel. Topics Signal Process.}, vol.~3, no.~2, pp. 253--265,
  2009.

\bibitem{longbotham1989theory}
H.~G. Longbotham and A.~C. Bovik, ``Theory of order statistic filters and their
  relationship to linear fir filters,'' \emph{IEEE Trans. Acoust., Speech,
  Signal Process.}, vol.~37, no.~2, pp. 275--287, 1989.

\bibitem{hosu2017konstanz}
V.~Hosu, F.~Hahn, M.~Jenadeleh, H.~Lin, H.~Men, T.~Szir{\'a}nyi, S.~Li, and
  D.~Saupe, ``The {K}onstanz natural video database ({KoNViD-1k}),'' in
  \emph{Proc. Int. Conf. Qual. Multimedia Exper. (QoMEX)}, 2017, pp. 1--6.

\bibitem{sinno2018large}
Z.~Sinno and A.~C. Bovik, ``Large-scale study of perceptual video quality,''
  \emph{IEEE Trans. Image Process.}, vol.~28, no.~2, pp. 612--627, 2018.

\bibitem{bampis2018spatiotemporal}
C.~G. Bampis, Z.~Li, and A.~C. Bovik, ``Spatiotemporal feature integration and
  model fusion for full reference video quality assessment,'' \emph{IEEE Trans.
  Circuits Syst. Video Technol.}, 2018.

\bibitem{pei2015image}
S.-C. Pei and L.-H. Chen, ``Image quality assessment using human visual {DOG}
  model fused with random forest,'' \emph{IEEE Trans. Image Process.}, vol.~24,
  no.~11, pp. 3282--3292, 2015.

\bibitem{xue2014blind}
W.~Xue, X.~Mou, L.~Zhang, A.~C. Bovik, and X.~Feng, ``Blind image quality
  assessment using joint statistics of gradient magnitude and {L}aplacian
  features,'' \emph{IEEE Trans. Image Process.}, vol.~23, no.~11, pp.
  4850--4862, 2014.

\bibitem{kundu2017no}
D.~Kundu, D.~Ghadiyaram, A.~C. Bovik, and B.~L. Evans, ``No-reference quality
  assessment of tone-mapped {HDR} pictures,'' \emph{IEEE Trans. Image
  Process.}, vol.~26, no.~6, pp. 2957--2971, 2017.

\bibitem{ye2012unsupervised}
P.~Ye, J.~Kumar, L.~Kang, and D.~Doermann, ``Unsupervised feature learning
  framework for no-reference image quality assessment,'' in \emph{Proc. IEEE
  Conf. Comput. Vis. Pattern Recognit.}, 2012, pp. 1098--1105.

\end{thebibliography}

\end{document}